\documentstyle[emulateapj, apjfonts, psfig]{article}

\def\etal{{et~al.}}

\def\amin{\ifmmode^{\prime}\else$^{\prime}$\fi}
\def\asec{\ifmmode^{\prime\prime}\else$^{\prime\prime}$\fi}

\def\simgt{\lower.5ex\hbox{$\; \buildrel > \over \sim \;$}}
\def\simlt{\lower.5ex\hbox{$\; \buildrel < \over \sim \;$}}

\newcommand\rcw{\hbox{RCW\hspace{1.5pt}103}}
\newcommand\src{\hbox{1E\hspace{1.5pt}161348$-$5055}}

\newcommand\Einstein{{\it Einstein}}
\newcommand\einstein{{\it Einstein}}
\newcommand\ASCA{{\it ASCA}}
\newcommand\asca{{\it ASCA}}
\newcommand\ROSAT{{\it ROSAT}}
\newcommand\rosat{{\it ROSAT}}

\slugcomment{Draft of Jan 19 1999}

\lefthead{Gotthelf, Petre \& Vasisht}

\righthead{X-ray Variability from \rcw} 

\begin{document}

\title{X-ray Variability from the Compact Source in the Supernova Remnant \rcw}

\author{E. V. Gotthelf$^1$, R. Petre$^2$ \&\ G. Vasisht$^3$}
\altaffiltext{1}{Columbia Astrophysics Laboratory, Columbia University, 550 West 120$^{th}$ Street, New York, NY 10027, USA; evg@astro.columbia.edu} 
\altaffiltext{2}{NASA/Goddard Space Flight Center, Code 662, Greenbelt MD, 20771, USA; petre@gsfc.nasa.gov}
\altaffiltext{3}{ Jet Propulsion Laboratory, California  Institute of Technology, 4800 Oak Grove Drive, Pasadena, CA, 91109, USA; gv@astro.caltech.edu}

\begin{abstract}

A new \ASCA\ observation of \src, the central compact X-ray source in
the supernova remnant \rcw, reveals an order-of-magnitude decrease in
its $3-10$ keV flux since the previous \ASCA\ measurement four years
earlier. This result is hard to reconcile with suggestions that the
bulk of the emission is simple quasi-blackbody, cooling radiation from
an isolated neutron star.  Furthermore, archived \einstein\ and
\rosat\ datasets spanning 18 years confirm that this source manifests
long-term variability, to a lesser degree.  This provides a natural
explanation for difficulties encountered in reproducing the original
\einstein\ detection of \src. Spectra from the new data are consistent
with no significant spectral change despite the decline in
luminosity. We find no evidence for a pulsed component in any of the
data sets, with a best upper limit on the pulsed modulation of 13
percent. We discuss the phenomenology of this remarkable source.

\end{abstract}


\section{Introduction}

Recent spectro-imaging X-ray  observations of central compact sources
in supernova remnants (SNRs) challenge earlier notions that most young
neutron stars (NSs) evolve in a manner similar to the prototypical
Crab pulsar (Gotthelf 1998). In fact, the latest compilations
shows that most of such associated objects manifest properties
distinct from those of the Crab-like systems.

Based on observational grounds alone, three classes of NSs in SNRs
are known whose flux is dominated by their X-ray emission; these
include the X-ray pulsars with anomalously slow rotation (with periods
in the range of $5-12$ s) and steep ($\Gamma \simgt 3$) power-law spectra
(Gregory \& Fahlman 1980; Gotthelf \& Vasisht 1998 and refs. therein),
the soft gamma-ray repeaters (Cline \etal\ 1982; Kulkarni \etal\ 1994;
Vasisht \etal\ 1994), and a population of radio-quiet NSs in remnants
(Caraveo \etal\ 1996; Petre \etal\ 1996; Mereghetti \etal\ 1996).  The
above objects are linked by their apparent radio-quiet nature, and
taken collectively, may help further reconcile the NS birth rate with
the observed SNR census. In this study we focus on the enigmatic X-ray
source \src\ in the SNR \rcw, for which no clear interpretation yet
exists within the above taxonomic framework.

The \einstein\ X-ray source \src\ lies near the projected center of
the bright, young ($\sim 2\times 10^3$ yrs; Carter \etal\ 1997)
Galactic shell-type SNR \rcw\ (G332.4-0.4) and has been proposed as
the first example of an isolated, cooling NS (Tuohy \& Garmire
1980). It was discovered using the high resolution imager (HRI) but
went unseen by a prior \einstein\ IPC observation and a subsequent
\rosat\ PSPC one, supposedly due to the poorer spatial resolution of
these instruments. Surprisingly, an initial observation with the
\rosat\ HRI also failed to detect the source; this was attributed to
the reduced HRI sensitivity of the 10' off-axis pointing (Becker
1993). Finally, a 1993 \asca\ observation re-discovered this elusive
object (Gotthelf, Petre \& Hwang 1997, GPH97 herein), but its spectral
characteristics were found to be incompatible with a simple cooling
NS model. This re-detection has been confirmed by more
recent, on-axis, \rosat\ HRI observations.

Herein, we present the results of our follow-up (Sep 1997)
\asca\ observation of \src. In the same field lies the recently
discovered 69 ms pulsar AX~J161730-505505 (Torii \etal\ 1998), whose
analysis is presented separately (Torii \etal\ 1999). While both
sources are detected again, the flux from \src\ has declined
significantly since the previous \asca\ measurement.  We discuss some
implications of this large flux variability on the nature of \src.

\section{Observations and Analysis}

A day-long follow-up observation with the \asca\ Observatory (Tanaka
\etal\ 1994) of \rcw\ was carried out on 1997 September 4. Data were
acquired with both the solid-state (SISs) and gas scintillation
spectrometers (GISs).  The essential properties of these instruments
are qualified in GPH97.  The SIS data were acquired in 1-CCD BRIGHT
mode with \src\ placed as close to the mean SIS telescope bore-sight
as was practical, to minimize vignetting losses.  The GIS data were
collected in the highest time-resolution mode ($0.5 \ \rm{ms}$ or
$0.064 \ \rm{ms}$, depending on the telemetry mode), with reduced
spectral binning of $\sim 12$ eV per PHA channel.  The effective,
filtered observation time is $58(49)$~ks for each GIS(SIS) sensor. The
new data were reduced and analyzed with the same methodology as in
GPH97.

\section{Results}

We compared images of \rcw\ from our new observation with the ASCA
images obtained four years earlier. The flux-corrected GIS images from
the two epochs, restricted to the hard energy band-pass ($3-10$ keV),
are displayed in figure 1 using an identical intensity scale. Both reveal
a pair of distinct features, each having a spatial distribution
consistent with that of a point source; one at the position of \src,
and the other at the position of the 69 ms X-ray pulsar
AX~J161730-505505 (due north); the flux of the latter has evidently
remained constant (see Torii \etal\ 1999 for details). However, we
estimate that \src\ has dimmed a factor $\simeq 12$ in the hard
band, after the diffuse flux from the remnant has been taken into
account using the following method.

The soft X-ray flux (below 2 keV) is dominated by steady thermal
emission from shock-heated gas in the remnant.  The contribution of
this component to the hard-band images is estimated using the
soft-band images. The latter provides a good model for the spatial
distribution of the surrounding shell on arcminute scales in the
$3-10$ keV range. The soft-band contribution from the shell was
renormalized to the hard-band, and subtracted from the flux calibrated
hard-band image to extract the flux contribution from the source
alone.  For the comparison, the new data were rebinned by a factor of
4 ($\sim 1^{\prime} \times 1^{\prime}$ pixels) to match the binning
used with the earlier observation.  The longer exposure of the second
observation results in increased sensitivity to \src, however, this
gain is offset by its location at a greater off-axis angle (as is
evident by the asymmetrical PSF) relative to the first observation. An
equivalent analysis of the SIS data reproduces the variability seen in
the GIS hard band.

\subsection{Spectroscopy}

We analyzed the spectrum from the new observation using the same
approach presented in GPH97. To maximize the sensitivity, we
simultaneously fit the spectra from all four \asca\ detectors. We
restricted our SIS spectral fits to $> 1.2$ keV as the calibration at
the lower energies has become less reliable over time.  The
improved viewing geometry over the previous observation, coupled with
the spectral stacking (four detectors), made it possible to
measure the spectrum, despite the lower source flux (only one SIS
data set was used in the earlier spectral analysis).  The resulting
fits to simple models (Table I) are consistent with those inferred
for the first observation. Thus, while the flux has decreased by an
order of magnitude, the spectrum appears essentially unchanged.

\subsection{The Long-term Light Curve}

To investigate its long-term flux behavior, we constructed a light
curve of \src\ which spans 18 years ($1979-1997$), using 10 available
archival observations. For each observation,
we extracted background-subtracted countrates or 3$\sigma$
upper limits. These rates are then used to estimate the flux in
a given energy band. For lack of knowledge to the contrary, we assume
that the spectral shape is invariant in time and modeled by a
blackbody whose parameters are given in Table 1. The best-fit
power-law model is unphysical at the softer X-ray energies and thus the
blackbody model is preferred for this comparison. We folded the latter
model through the spectral response function of each instrument using
the XSPEC spectral fitting package and inferred the source flux for
each observation in a fiducial $0.5-2$ keV energy band. The results,
listed in Table 2 and plotted in Fig. 2, confirm a dramatic flux
change between the \asca\ observations, and suggest that \src\ is
variable, to a lesser extent, amongst the other observations.

We present this source flux comparison among instruments with some
caution, as these can be potentially unreliable. The different energy
bands, flux calibrations, point response functions, and background
contamination can produce large uncertainties in the derived
fluxes. When extrapolating, the relative count rates are very
sensitive to the instrumental energy band, along with the assumed
${N_H}$ and emission model.  However, our fundamental result stands
regardless of the aforementioned caveats: the \asca\ data alone, and
to a lesser extent the \rosat\ data, establish the fact that \src\
varied throughout the time it has been observed.

\subsection{The Short-term Variability}

The excursions in flux noted from observations separated by
months or years suggest that variability might be present on shorter
time-scales.  We searched the day-long \asca\ observations for
hour-scale temporal variability.  In addition, we
examined the behavior on a timescale of a few days using
the August, 1995, \rosat\ HRI observation, with a net exposure
time of 50 ks spread over a week.  In neither case did we find evidence of
variability greater than the photon statistic limit of 10 percent.

We searched the new \asca\ GIS data for a coherent pulsed signal,
as in GPH97. No significant periodicity was
found in the period space between 10 ms - 1000 s.  The upper limit
on the pulsed fraction for this data set is $\sim$ 23\%, compared
with the 13\% limit of GPH97. The data show no clear evidence for 
accretion noise such as redness in the power spectrum. 
The data were searched for X-ray bursts and other
anomalies in the light curve, but none were found.

\section{Discussion}

A number of hypotheses have been advanced to explain the nature of
\src. Upon discovery, this source was proposed as an isolated neutron
star emitting blackbody radiation (Tuohy \& Garmire 1980). Further
optical and radio observations (e.g. Tuohy \etal\ 1983; Dickel \etal\
1996) have failed to identify a counterpart, thereby bolstering this
interpretation. The observations of GPH97 showed that the point source
could be described as a hot blackbody of $kT \simeq 0.6$ keV and a
$0.5-10$ keV flux of $6.5\times 10^{-12}$ erg cm$^{-2}$ s$^{-1}$;
therefore a luminosity of $L_X \simeq 10^{34}$ erg s$^{-1}$ (at 3.3
kpc) and an effective emitting area of $1$ km$^2$, or $\sim 0.1$\% the
surface area of a NS. This corresponds to a rather small hot-spot on a
stellar surface, which is in turn surprising since the source shows no
rotational modulation (down to $\sim 13$\%). Also, the inferred
temperature is too high for an object of age few $\times 10^3$ yrs
(but see below).

Heyl \& Hernquist (1998) recently attempted to salvage the cooling NS
model by invoking an ultra-magnetized star ($B_s \sim 10^{15}$ G) with
an accreted hydrogen atmosphere (Page 1997, for a
review). This combination can enhance the cooling flux as well as
shift the emission bluewards (Chabrier, Potekhin \& Yakov 1997) so
that it effectively mimics a hot blackbody in the \ASCA\ spectrum.
However, it is hard for cooling models to address the issue of
variability or the flux decimation observed between the \ASCA\ epochs,
unless there is a source for rejuvenating the heating of the NS
interior. Such a source could be the super-strong magnetic field,
implicit in the above model. Episodic rearrangement of the field
(Thompson \& Duncan 1996) in the stellar interior could provide the
energy to impulsively heat the core.

The stellar surface would re-adjust to reflect the internal heating on
a short thermal timescale of a few months.  Although heating of the NS
is viable in this scenario, the rapid cooling on a timescale of a few
years, observed between the \ASCA\ epochs, cannot be explained without
some very ``exotic'' cooling process
.  In addition, a factor of 10 variability in the hard ($3 - 10$ keV)
band would result in a downward shift of the effective temperature by
a factor 1.4, which should have been detected. On these grounds, we
reject the hypothesis that the observed X-ray emission from \src\ is
simple cooling radiation.

An ultra-magnetized NS is also a leading model for the anomalous X-ray
pulsars (AXPs), an example of which is the $\simlt 2,000$ yr-old, 12-s
X-ray pulsar in the remnant Kes 73 (Vasisht \& Gotthelf 1997).  GPH97
compared \src\ to the latter on the basis of similar spectral
characteristics.  And at least two AXPs are reported to vary
significantly in flux, by as much as a factor of five (1E~1048.1-593;
Oosterbroek \etal\ 1998). While \src\ shows some properties that are
tantalizingly similar to those of the AXPs, the lack of observed
strong pulsations ($\sim 30\%$ modulation for the AXPs) is notably
amiss, particularly as a large magnetic field should result in highly
anisotropic surface emission. Gravitational defocusing and/or
unfavorable viewing geometry, however, might account for the lack of
observed pulsations.

Alternatively, the variability can be indicative of an accreting
compact object.  Popov (1997) suggests that \src\ is an old accreting
NS with a low magnetic field and long spin period ($\sim 10^3$ s), the
by-product of a disrupted binary and not of the same age as \rcw.
This proposition is bolstered by the discovery of the nearby 69 ms
pulsar AX~J161730-505505 (spindown age of $8100$ yrs), located outside
the remnant shell (Torii \etal\ 1998). Several arguments, including
those based on the pulsar's implied velocity and lack of wind nebula,
and the symmetry of the SNR, however, make an association unlikely
(Kaspi \etal\ 1998, Torii \etal\ 1999).

Finally, there exists the possibility that the source is an isolated
stellar-mass black hole (BH), accreting from the surrounding medium or
from supernova ejecta fall-back. Brown \& Bethe (1994) have discussed
scenarios in which a massive progenitor explodes as a supernova and
then evolves into a BH of several solar masses after accreting
captured ejecta.  Such a scenario is clearly applicable to the source
in \rcw. Temporal variability and lack of pulsed emission are the
natural consequences in such a model.
                  
An accretion process around a BH almost inevitably involves rotating
gas flows. Popov (1997) has dismissed the possibility of a few
stellar-mass BH in \rcw\ based on the small implied emitting area
($\sim 1$ km$^2$) of an equivalent blackbody radiator (see \S 4).
This argument would certainly apply for the
case of the standard optically-thick, thin-disk model. However,
low-efficiency solutions can exist for accretion flows (especially at
low $\dot M$) around BHs, in which most of the viscously generated
thermal energy is advected into the BH. 
Below a critical mass accretion rate of $\sim 0.1 \dot
M_{Edd}$ ($\dot M_{Edd}$ is the Eddington rate) accretion flows turn
advection dominated (or ADAF), and the observed luminosity in \src\
would suggest an accretion rate of $10^{-(2.5-3.0)} \dot M_{Edd}$~ (or
$\sim 10^{-10}$ $M_\odot$ yr$^{-1}$) 
(cf ADAF models summarized in Narayan \etal\ 1998).
At this rate, the BH would accrete $\sim 10^{-7}$ $M_\odot$ of matter,
a small fraction of the mass of supernova ejecta, at the sustained
present rate over its lifetime of $\sim 10^3$ yr.
Within the framework of the above arguments, it is possible that flow around
\src\ could be detected as a faint optical source (V $\simgt$ 22 after
accounting for visual extinction) or a 10-100 $\mu$Jy (1 GHz) radio source.

\section{Conclusions}

The variability of the X-ray emission from the compact source in \rcw\
leaves little room for a conventional cooling NS origin. Instead, an
accretion scenario may be considered, although the relatively low
luminosity, lack of optical counterpart, and young age are
inconsistent with a typical accreting NS binary. Accretion from a very
low mass ($\le$0.1 M$_{\odot}$) companion (Mereghetti \etal\ 1996;
Baykal \& Swank 1996) or a fossil disk around a solitary NS (van
Paradijs \etal\ 1995), however, is not ruled out. We suggest that
within the context of inefficient accretion (such as advection
dominated flows), a stellar mass black hole is a viable
possibility. We reiterate, however, that the spectral characteristics
of \src\ are remarkably similar to those of the AXPs, which are
suspected to be ultra-magnetized NSs and thought to be powered by
magnetic field decay rather than rotational braking (Thompson \&
Duncan 1996; Vasisht \& Gotthelf 1997).

Independent of the above phenomenology, the properties of \src\ add to
the view that young collapsed stars can follow an evolutionary
scenario quite distinct from those of Crab-like pulsars. The property
of being radio quiet is common to all AXPs, SGRs and to \src\
(Gotthelf 1998; for possible physical
mechanisms see 
Baring \& Harding 1998). It is
likely that they all share a common heritage and may prove to be part
of an evolutionary sequence.

\begin{acknowledgements}

This work uses data made available from the HEASARC public archive at
GSFC. G.V. thanks J Heyl for discussions.

\end{acknowledgements}

\clearpage

\begin{deluxetable}{lccc}
\footnotesize
\tablewidth{0pt}
\tablecaption{Fits to \ASCA\ SIS Spectrum \vfill
\label{tbl-1}}
\tablehead{
 \colhead{\hfil Model$^a$ \hfil} & \colhead{$\chi^2 (\rm{DoF})$} & \colhead{kT or $\Gamma$$^b$} &
 \colhead{kT or $\Gamma$$^b$}  \nl
 \multispan3{ \dotfill 1997 \dotfill}  & \colhead{1993}
}
\startdata
 Diffuse Surroundings NEI  & 390 (227)  & 0.3 & 0.3  \nl
 Source NEI  & 417 (211) &  -  &   -    \nl
 NEI + BB    & 260 (209) & $0.63_{0.57}^{0.71}$ & $0.56_{0.53}^{0.59}$ \nl
 NEI + PL    & 238 (209) & $3.9_{3.6}^{4.1}   $ & $3.2_{3.0}^{3.4}   $ \nl
 NEI + BREM  & 247 (209) & $1.43_{1.20}^{1.76}$ & $1.6_{1.4}^{1.8}   $ \nl
\enddata
\tablenotetext{a} {See GPH97 for details.}
\tablenotetext{b} { $\Gamma$ is the photon index; kT in units of keV. $\rm{N_H}$ is fixed to the best fit values of $7.3 \times 10^{21}$ cm$^{2}$ (1997) and $6.8 \times 10^{21}$ cm$^{2}$ (1993).}
\end{deluxetable}

\begin{deluxetable}{l c c c c}
\tablewidth{0pt}
\footnotesize
\tablecaption{Log of X-ray Imaging Observations of \rcw\ \vfill
\label{tab:var-short}} 
\tablehead{
\colhead{Mission/Sensor} & \colhead{Obs. Date} &
\colhead{Expo.} & \colhead{Cnt rate$^a$} & 
\colhead{.5--2 keV flux$^b$ }  \\ 
                                  & \colhead{(yy/mm/dd)}      &
\colhead{(ks)}   & \colhead{(cps)}     & 
\colhead{(10$^{-13}$ cgs)}
}
\startdata
\Einstein\ / IPC     & 79/02/26 & 2.6      & $<$40         &$<$5.1           \nl
\Einstein\ / HRI     & 79/09/14 & 3.1      & 3.8$\pm$0.7   &4.4$\pm$0.7      \nl
\ROSAT\	   / HRI     & 91/02/13 & 5.1      & $<$7.5        &$<$3.2           \nl
\ROSAT\	   / PSPC    & 91/02/27 & 37.2     & 7.7$\pm$2.3   &1.1$\pm$0.3      \nl
\ASCA\	   / SIS     & 93/08/17 & 47.7     & 50$\pm$10$^c$ &55$\pm$11        \nl
\ROSAT\	   / HRI     & 94/02/26 & 1.1      & 13$\pm$4      &6.0$\pm$1.7      \nl
\ROSAT\	   / HRI     & 94/08/02 & 4.4      & 9$\pm$2.6     &3.8$\pm$1.2      \nl
\ROSAT\	   / HRI     & 95/03/08 & 48.1     & 8.4$\pm$0.7   &3.6$\pm$0.4      \nl
\ROSAT\	   / HRI     & 95/08/18 & 8.1      & 7.4$\pm$1.6   &3.1$\pm$0.6      \nl  
\ASCA\	   / SIS     & 97/09/04 & 58.6     & 4$\pm$1$^c$   &6$\pm$1          \nl
\enddata
\tablenotetext{a} {Spectral fit band-pass: $0.2-4.5$ keV \Einstein; $0.2-2.4$ keV \rosat; $1.2 - 10$ keV \ASCA.}
\tablenotetext{b} {Flux at earth in ergs cm$^{-2}$ s$^{-1}$, assuming best fit 
blackbody ($k\rm{T}$=0.63 keV) and N$_H$ (7.3$\times$10$^{21}$ cm$^{-2}$) from the second \asca\ observation. Upper limits are $3\sigma$.}
\end{deluxetable}


\begin{figure}[here]
\centerline{
\hfill\hfill
\psfig{figure=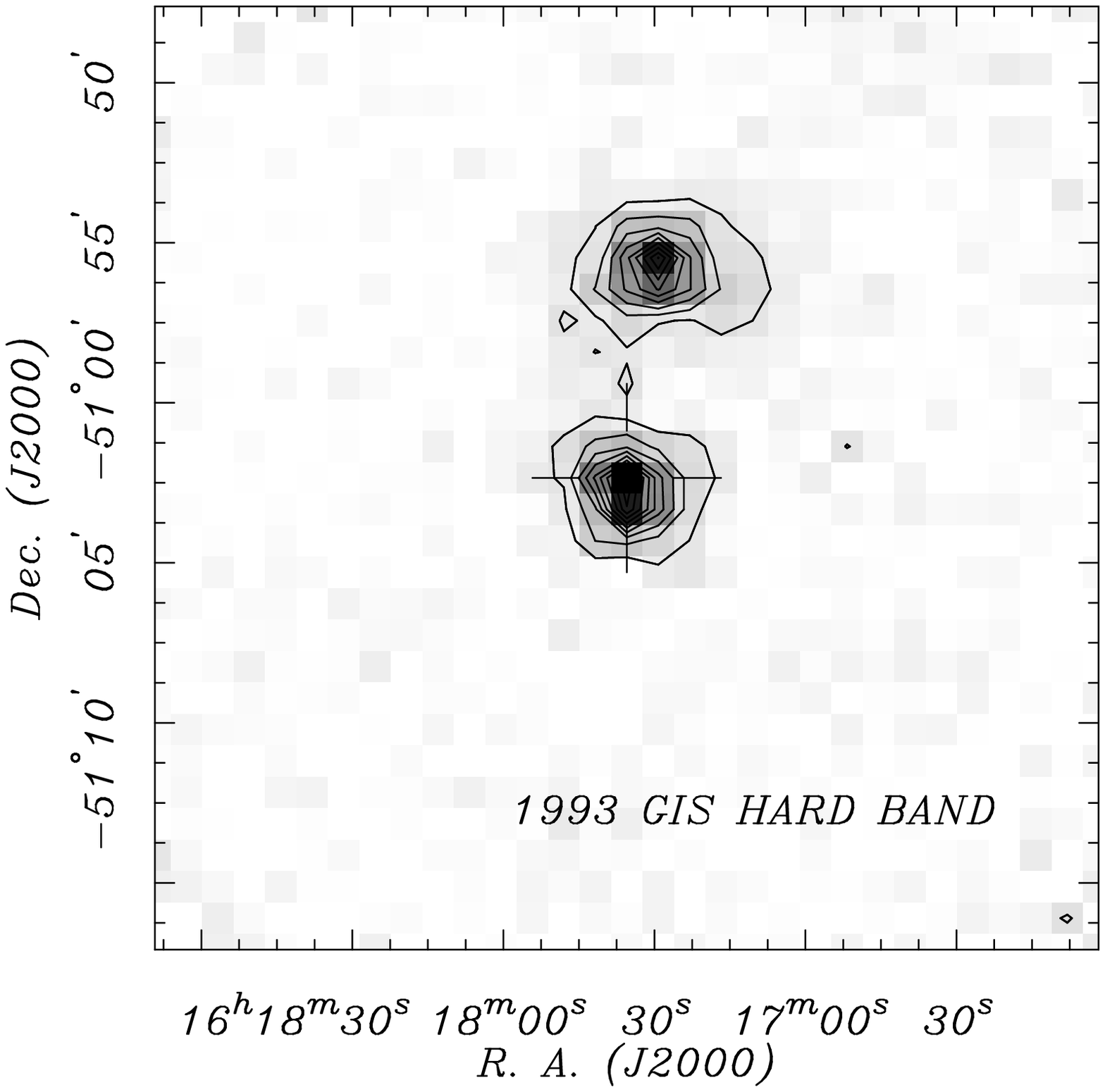,height=9cm,angle=270.0,bbllx=25bp,bblly=25bp,bburx=587bp,bbury=557bp,clip=}
\psfig{figure=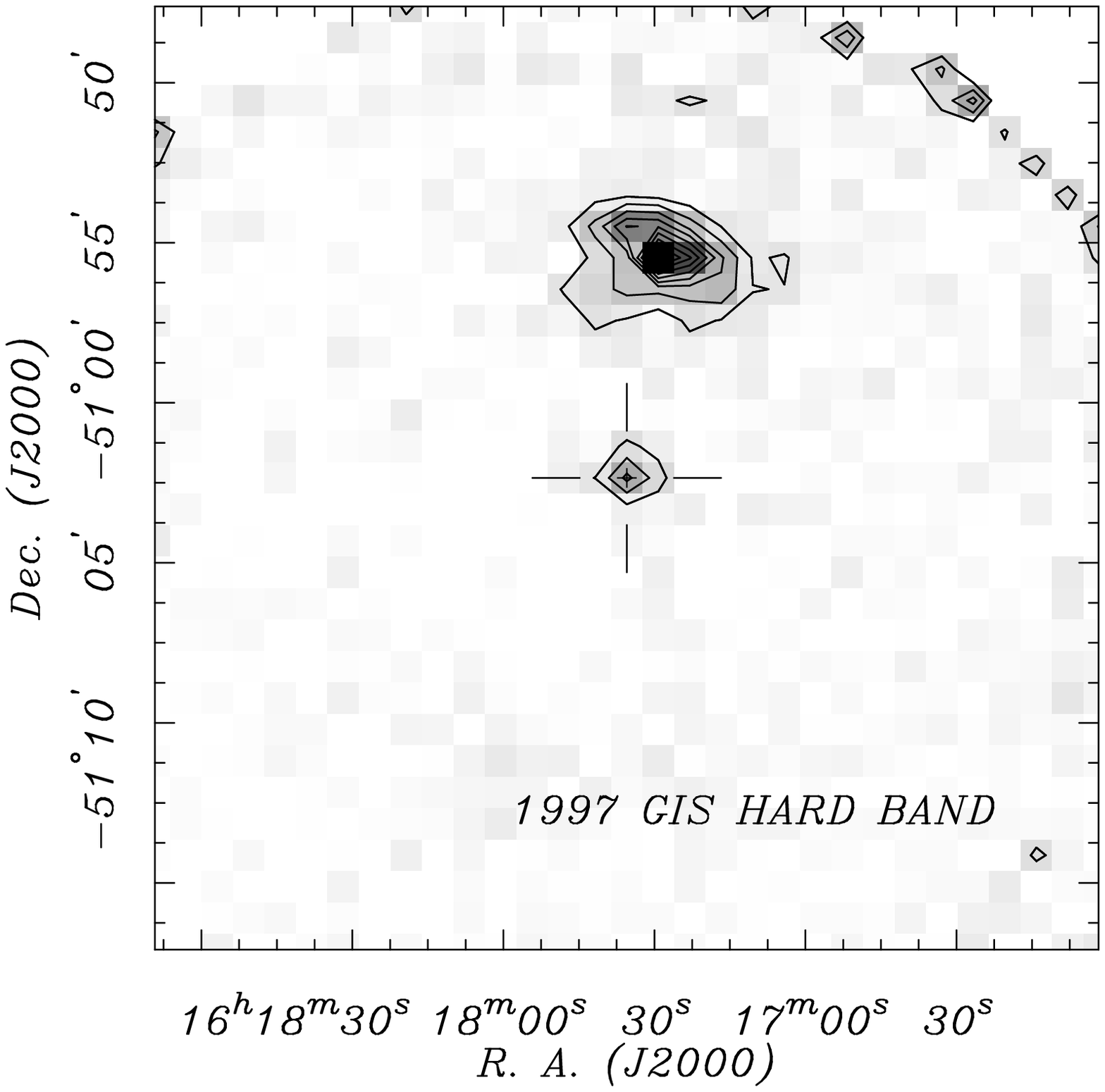,height=9cm,angle=270.0,bbllx=25bp,bblly=25bp,bburx=587bp,bbury=557bp,clip=}
\hfill\hfill
}

\caption{\footnotesize ASCA GIS images of \src\ (cross), the compact
point source in \rcw, taken 4 years apart (Left -- 1993; Right --
1997). At the top of the picture is the 69 ms X-ray pulsar
AX~J161730-505505. These background subtracted hard energy band-pass
($3 - 10$ keV) images are flux corrected and plotted with the same
scaling. The central source in \rcw, 1E~161348-5055, has dimmed by a
factor of $\sim 12$, while the pulsar flux remained relatively
constant. The soft nebula flux from the shell of \rcw\ has
been subtracted from these images.  Contours along the edge of GIS in
the 1997 observation (upper right corner) are not significant.  }

%

\vspace{1.0truein}

\centerline{
\psfig{figure=rcw103_gpv98_fig2.ps,height=3.0truein,angle=270.0,clip=}
}
\caption{\footnotesize The long term light curve of \src\ in the $0.5-2.0$ keV
energy band-pass. Symbols represent: \einstein\ IPC (open circle);
\einstein\ HRI (filled circle); \ROSAT\ PSPC (open square), HRI
(filled squares); and \asca\ (star). Error bars are one sigma.  The
dashed vertical lines indicate the model dependent uncertainty in
the \asca\ flux when extrapolated to the $0.5 - 2$ keV band-pass.}

\end{figure}

\end{document}